# The Coexistence of Spinon Superconductivity and $d_{x^2-y^2}$ Superconductivity in Bi-2212 from Tunneling Measurements.

A. Mourachkine

*Université Libre de Bruxelles, Service de Physique des Solides, CP233, Boulevard du Triomphe, B-1050 Brussels, Belgium*
(Received January 1999)

We present electron-tunneling spectroscopy of slightly overdoped $Bi_2Sr_2CaCu_2O_{8+x}$ (Bi2212) single crystals at low temperature using a break-junction technique. Below $T_c$, we observe, at least, two different superconducting gaps. We show that (i) the largest superconducting gap occurs due to pairing of spinons; (ii) the second superconducting gap has the $d_{x^2-y^2}$ symmetry; (iii) the $d_{x^2-y^2}$ gap is smaller than the spinon gap, however, predominant and occurs due to pairing of fermionic excitations carrying charge; (iv) the spinon gap has either a s-wave or (s+d) mixed symmetry with the maximum located in $(\pi, \pi)$ direction on the Fermi surface. It is possible that there exists the third superconducting gap having a g-wave symmetry, which is approximately twice smaller in magnitude than the d-wave gap. Thus, there are two different types of superconductivity and two different types of carriers in Bi2212.

Since the discovery of high-temperature superconductors (HTSC) in 1986 by Bednorz and Müller [1] there is no complete consensus on the mechanism of superconductivity (SC) in copper-oxides. The definitive confirmation of the predominant $d_{x^2-y^2}$ (hereafter, d-wave) character of the SC in the cuprates is a great advance in recent years [2]. However, the interpretation of some experimental data shows the presence of a s-wave component in cuprate superconductors [3]. Emery, Kivelson and Zachar [4] have presented a theory of spinon SC in HTSC. Spinons are neutral fermions which occur in one-dimensional (1D) physics. However, up to now there is no direct evidence of the spinon SC in cuprates. They emphasize that any theory which involves pairing of real-space particles in $CuO_2$ planes are *a priori* implausible due to the strong short-range Coulomb repulsion between two carriers. Indeed, it is most likely that quasiparticles in cuprates at low temperatures are the bound state of excitations. Mihailovic and Müller [5] have presented strong indications for simultaneous presence of polaronic and fermionic carriers in mesoscopic areas in the $CuO_2$ planes below $T_c$.

In this Letter, we present direct measurements of the density of states by electron-tunneling spectroscopy on overdoped $Bi_2Sr_2CaCu_2O_{8+x}$ (Bi2212) single crystals at low temperature using a break-junction technique. Below $T_c$, we observe, at least, two different SC gaps. We show that (i) the largest SC gap occurs due to pairing of spinons; (ii) the second SC gap has the d-wave symmetry; (iii) the d-wave gap is smaller than the spinon gap, however, predominant and occurs due to pairing of fermionic excitations carrying charge; (iv) the spinon gap has either a s-wave or (s+d) mixed symmetry with the maximum located in $(\pi, \pi)$ direction on the Fermi surface. It is possible that there exists the third SC gap having a g-wave symmetry, which is approximately twice smaller in magnitude than the d-wave gap. To our knowledge, it is for the first time the direct evidence of the spinon SC is presented in the literature, which coexists with another type of the SC with the d-wave gap which is predominant in Bi2212.

The tunneling spectroscopy has played the crucial role in the verification of the BCS theory [6], which is particularly sensitive to the density of state (DOS) near the Fermi level ($E_F$) and, thus, is capable of detecting *any* gap in the quasiparticle excitation spectrum at $E_F$ [6]. In addition to this, it has a very high energy resolution (less than $k_B T$ for a superconductor-insulator-superconductor (SIS) junction) [6]. The B-J technique can detect the Josephson current, hence, can distinguish between a SC and non-SC gap.

The single crystals of Bi2212 were grown using a self-flux method and then mechanically separated from the flux in $Al_2O_3$ or $ZrO_2$ crucibles [7]. The dimensions of the samples are typically $3 \times 1 \times 0.1$ mm$^3$. The chemical composition of the Bi2212 phase corresponds to the formula $Bi_2Sr_{1.9}CaCu_{1.8}O_{8+x}$ in overdoped crystals as measured by energy dispersive X-ray fluorescence (EDAX). The single crystals were checked out to assure that they are in the overdoped phase. The crystallographic *a, b, c* values of the overdoped single crystals are of 5.41 Å, 5.50 Å and 30.81 Å, respectively. The $T_c$ value was determined by either dc-magnetization or by four-contacts method yielding $T_c$ = 87 - 90 K with the transition width $\Delta T_c \sim 1$ K. Experimental details of our B-J technique are given in Refs. 7 and 8.

Typical conductance curves *dI/dV(V)* and tunneling current-voltage *I(V)* characteristics for our SIS break junctions on Bi2212 single crystals can be found elsewhere [8, 9]. They exhibit the characteristic features of typical SIS junctions [10, 11]. The magnitude of a SC gap can, in fact, be derived directly from the tunneling spectrum. However, in the absence of a generally accepted model for the gap function and the DOS in HTSC, such a quantitative analysis is not straightforward [12]. Thus, in order to compare different spectra, we calculate the gap amplitude $2\Delta$ (in m*e*V) as a half spacing between the conductance peaks at $\pm 2\Delta$.

Figure 1 shows a set of normalized tunneling spectra measured at 14 K as a function of bias voltage on an overdoped Bi2212 single crystal with $T_c$ = 89.5 K. The spectra B, C, D and E were measured in sequence by changing the distance in a junction (by bending the flexible substrate). The spectra A and F were obtained separately. The variation of the magnitude of the tunneling gap between 23 and 32.5 meV is in a good agreement with angle-resolved tunneling data [13] which are presented in Fig. 2, from which one can see that the maximum of the tunneling gap is located in ($\pi$, $\pi$) direction on the Fermi surface. The minimum of the tunneling gap is located at (0, $\pi$). Consequently, the spectra A and F shown in Fig. 1 were detected in (or close to) ($\pi$, $\pi$) and (0, $\pi$) directions, respectively. By other words, Figure 1 presents the angle dependence of the tunneling gap between ($\pi$, $\pi$) and (0, $\pi$) directions on the Fermi surface. By using the angular dependence of the gap we present also in Fig. 2 the angular dependence of the magnitude of the Josephson current for the data shown in Fig. 1. It is clear from Fig. 2 that the SC gap which corresponds to the maximum Josephson current has the $d_{x^2-y^2}$ symmetry. One can see also from the angular dependence of the Josephson current in Fig. 2 that, probably, there exists a small SC gap with the g-wave symmetry, which, most likely, corresponds to the sub-gap shown in Fig. 1.

We concentrate now on the absence of the Josephson current in $(\pi, \pi)$ direction on the Fermi surface. One can see in Fig. 1 that the magnitude of the Josephson current depends on the magnitude of the tunneling gap. It is noteworthy that such dependence of the value of the Josephson current on the magnitude of the tunneling gap is *typical* for each separate sample: as larger the magnitude of the tunneling gap as smaller the value of the Josephson current [14]. It is possible to explain the absence of the Josephson current on the spectrum A shown in Fig. 1 because of the normal resistance of the junction, $R_N$ is too high. However, a weak Josephson current was detected in junctions with $R_N$ higher than 120 kΩ. So, this is not the reason for the absence of the Josephson current. Consequently, a gap corresponding to the spectrum with the maximum magnitude of the tunneling gap (spectrum A in Fig. 1) has, probably, a non-SC origin.

Let's analyze the origin of the large tunneling gap (spectrum A in Fig. 1). Main conductance peaks of any normal-state gap will have the same bias positions in a superconductor-insulator-normal metal (SIN) junction and in a SIS junction. The bias positions of conductance peaks corresponding to any SC gap will be twice larger in a SIS junction than in a SIN junction. DeWilde, Miyakawa *et at.* [10] have performed tunneling measurements on the same set of optimally doped Bi2212 single crystals by STM (DeWilde) and by B-J technique (Miyakawa). In SIN junctions, the main peaks corresponding to the maximum value of the tunneling gap have been detected at ±37 meV [10]. In SIS junction, they have been observed at ±76 meV [10]. This fact points out that the large tunneling gap behaves like a SC gap (because 2×37 meV ≈ 76 meV). It is *striking* result that the large tunneling gap has the SC behavior and there is no Josephson current in tunneling spectra corresponding to this gap (spectra A in Fig. 1). There is only one explanation of this peculiar fact. The large tunneling gap is a SC gap which occurs due to pairing of chargeless fermions - spinons. In general, the SC occurs due to pairing of fermions, however, there is no requirements on charge. The Josephson current as any other current requires the charge transfer. However, if spin and charge are separated, for example, in the striped phase of the $CuO_2$ planes [15, 16] as in the 1D electron gas [4], then the large tunneling gap is a pure spin gap. The charge is carried by bosonic holons. There is no gap in the charge response. Consequently, this is the reason for the absence of the Josephson current. It is difficult to accept this fact since, up to now, we have only observed SCs with quasiparticles carrying charge. The Josephson current was the indicator of the SC. However, it is not the case for the spinon SC. The maximum of the spinon-SC gap (spectrum A in Fig. 1) is located in $(\pi, \pi)$ direction on the Fermi surface. The d-wave SC gap in $(0, \pi)$ direction occurs due to pairing of fermionic excitations carrying charge because the presence of the Josephson current.

We now turn to the symmetry of the spinon-SC gap. Figure 3 shows temperature dependencies of tunneling spectra of overdoped Bi2212. The curve A displays a temperature dependence of the spectrum F shown in Fig. 1 [17]. The curve B is a *typical* temperature dependence for a maximum tunneling gap or for a gap which is close to the maximum value [9, 11], *i. e.* this is a temperature dependence of the spinon-SC gap. From temperature dependencies A and B shown in Fig. 3, we conclude that there is one SC gap in ($\pi$, $\pi$) direction on the Fermi surface. It is logic since the magnitude of the d-wave SC gap is equal to zero in ($\pi$, $\pi$) direction. The second conclusion is that there are two SC gaps at (0, $\pi$). In two-band SC, the temperature dependence of one of the two gaps lies below the BCS temperature dependence [6]. This implies that the spinon-SC gap is non-zero at (0, $\pi$) on the Fermi surface. This points out that the spinon-SC gap has either a pure s-wave or mixed (s+d) symmetry. The inset in Fig. 3 shows schematically the two gaps on the Fermi surface in overdoped Bi2212. The maximum magnitude of the tunneling gap corresponds to the spinon-SC gap which is non-zero at (0, $\pi$) and ($\pi$, 0). If the g-wave SC gap is present in Bi2212, the maximums of the g-wave gap are located in ($\pi/4$, $3/4\pi$) and ($3/4\pi$, $\pi/4$) directions on the Fermi surface.

Let's compare the intensities of the spinon and d-wave gaps. The spectrum F in Fig. 1 mainly consists of the DOS corresponding to the d-wave gap. The spectrum A in Fig. 1 is the pure spinon-SC DOS. One can compare in Fig. 1 the intensities of the main peaks in the spectra A and F. In spite of the fact that the magnitude of the spinon gap is larger than the magnitude of the d-wave gap, the d-wave component (spectrum F) is more intense than the spinon-SC component (spectrum A). Indeed, the SC in Bi2212 has the predominant d-wave character.

Finally, we discuss some properties of the spinon SC and the SC with d-wave gap. Spinons are fermions, so it is most likely that the d-wave SC is a polaron SC [5]. Indeed, polarons are fermionic excitations carrying charge. If the d-wave SC is the polaron SC the next important question is to answer - what is 'the glue' for pairing of polarons. The mechanism of spinon pairing is described in Ref. 4. The spinon-SC gap will be detected only in experiments which are sensitive to spin. In Bi2212, the maximum magnitude of the tunneling gap (*i. e.* the spinon-SC gap) depends on hole concentration, $p$ in the $CuO_2$ planes linearly and has no correlations with the $T_c(p)$ dependence which is parabolic [18]. It is logic to associate the maximum magnitude of the d-wave SC gap with the $T_c(p)$ dependence. Hence, the $T_c$ is the characteristic temperature of the SC with the d-wave gap, and not of the spinon SC. In our study, the maximum *possible* magnitude of the d-wave gap can be equal to 23 meV for the overdoped Bi2212 single crystal with $T_c$ = 89.5 K. Then, we can estimate the maximum value of $2\Delta/k_BT_c$ for the SC with the d-wave gap: $2\Delta/k_BT_c$ 6. Since we know that the value 23 meV mainly consists of the d-wave gap we expect that the real $2\Delta/k_BT_c$ value will lies between $5 < 2\Delta/k_BT_c < 6$. There is no sense to calculate the $2\Delta/k_BT_c$ value for the spinon SC because they do not relate to each other.

In summary, we presented direct measurements of the density-of-state by tunneling spectroscopy on overdoped $Bi_2Sr_2CaCu_2O_{8+x}$ single crystals at low temperature using the break-junction technique. Below $T_c$, we observe, at least, two different superconducting gaps. We show that (i) the largest superconducting gap occurs due to pairing of spinons; (ii) the second superconducting gap has the $d_{x^2-y^2}$ symmetry; (iii) the $d_{x^2-y^2}$ gap is smaller than the spinon gap, however, predominant and occurs due to pairing of fermionic excitations carrying charge; (iv) the spinon gap has either a s-wave or (s+d) mixed symmetry with the maximum located in $(\pi, \pi)$ direction on the Fermi surface. It is possible that there exists the third superconducting gap having a g-wave symmetry, which is approximately twice smaller in magnitude than the d-wave gap. Thus, there are two different types of superconductivity and two different types of carriers in Bi2212.

I thank R. Deltour for support. This work is supported by PAI 4/10.

______________________________________________

Josephson current are not used in the presentation because of this requirement. However, it is not the case for break-junction technique.

FIGURE CAPTIONS:

FIG. 1. Tunneling spectra measured at 14 K on an overdoped Bi2212 single crystal with $T_c$ = 89.5 K. The conductance scale corresponds to spectrum A; the other spectra are shifted vertically for clarity by 1.2 units from each other. The spectra are normalized at -150 meV.

FIG. 2. Average measured tunneling gap at low temperature vs. angle on the Fermi surface (full line). The data are taken from Ref. 13. Normalized Josephson current, $I_J$ vs. angle for the data shown in Fig. 1 (dots). The dash line is a guide to the eye.

FIG. 3. Measured temperature dependencies of the quasiparticle DOS in Bi2212 single crystals. The dependence A (triangles) corresponds to the temperature dependence of the spectrum F shown in Fig. 1 [17]. The dependence B (dots) is a typical temperature dependence for a maximum SC gap [9, 11]. The solid line corresponds to the BSC temperature dependence. Inset: shapes of two SC gaps on the Fermi surface in overdoped Bi2212: black area (d-wave gap) and outlined area (spinon gap having either a s-wave or mixed (s+d) symmetry). The shapes of two gaps are shown schematically.

**(NOTE: the figures are in .ps format with poor quality for graphs)**

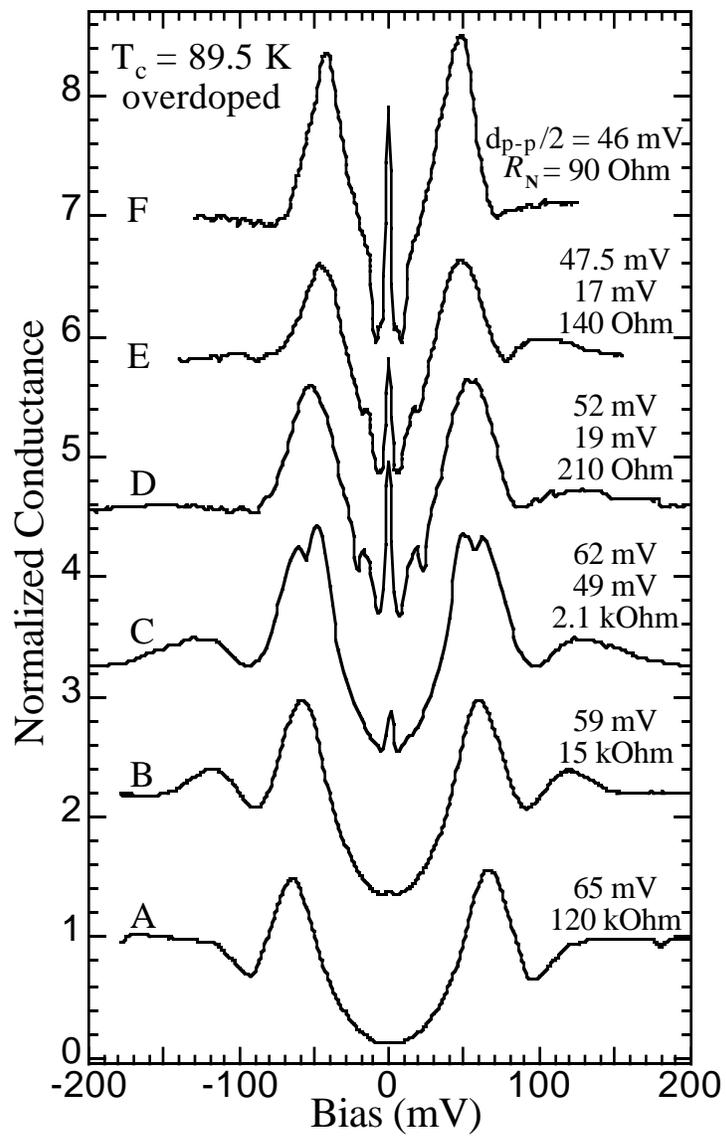

FIG. 1

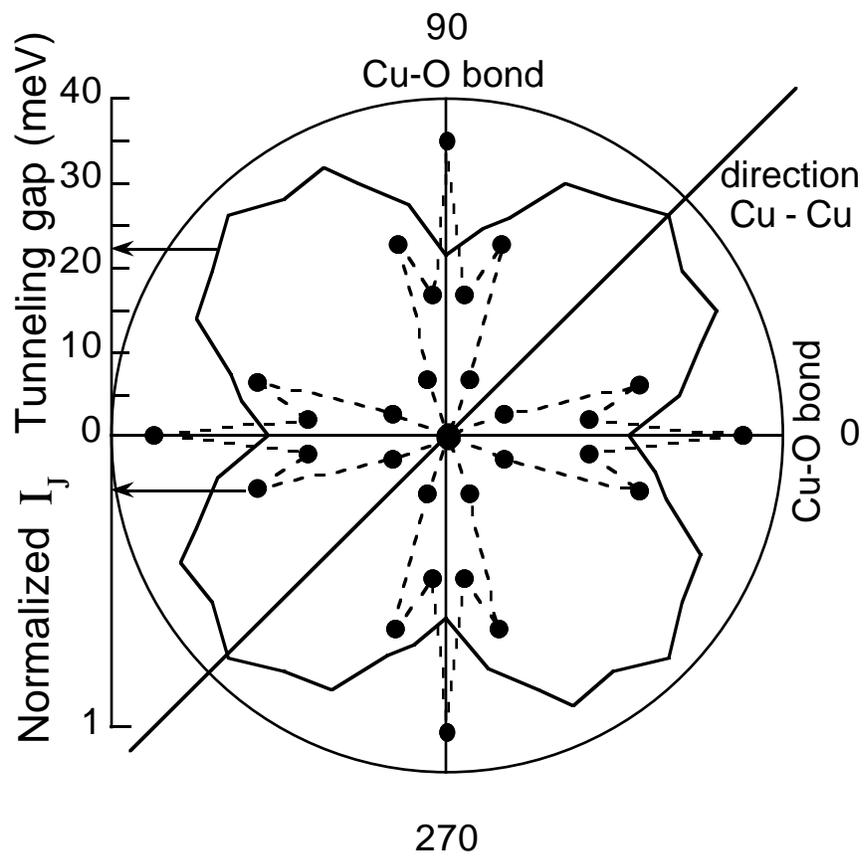

FIG. 2

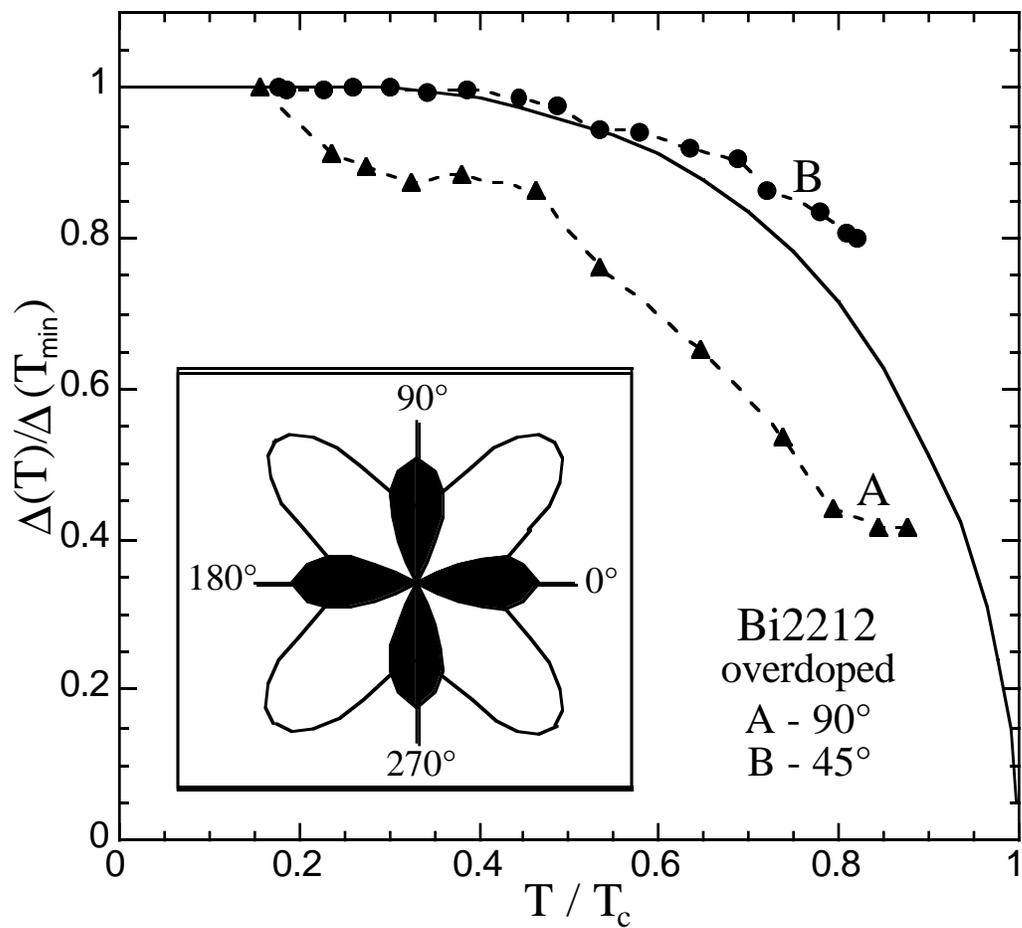

FIG. 3